\documentclass[conference]{IEEEtran}

\usepackage{graphicx}
\PassOptionsToPackage{hyphens}{url}
\usepackage{hyperref}
\usepackage{color}
\usepackage{flushend}

\usepackage{eqnarray,amsmath}
\usepackage{xspace}
\def\skipnoindent{\vskip0.1in\noindent}

\def\bend{~\rule[-0.015in]{0.075in}{0.075in}} 

\def\state#1{{\tt #1}}

\newcounter{exnum}

\newcounter{line}

\def\mc{\multicolumn}

\def\sensor#1#2{\hbox to 0.2in{#1:} #2}
\def\state#1{\text{#1}}
\def\invid#1{\hbox to 0.5in{#1:\hfill}} 
\def\state#1{{\tt #1}}

\def\hackfest{${\text S}^3$\xspace}
\def\hackfestend{${\text S}^3$\hskip-0.03in}
\def\waterdefense{WD\xspace}
\def\waterdefenseH{WDH\xspace}

\newcounter{attack}
\def\nextattack{\refstepcounter{attack}\theattack}

\IEEEoverridecommandlockouts

\begin{document}

\title{Assessing the Effectiveness of Attack Detection at a Hackfest on Industrial Control Systems \thanks{\noindent This work was supported by research  grant 9013102373 from the Ministry of Defense and NRF2014-NCR-NCR001-040 from  the National Research Foundation, Singapore.}}

\author{\IEEEauthorblockN{Sridhar Adepu and Aditya Mathur}
\footnote{XYZ}
\IEEEauthorblockA{iTrust, Center for Research in Cyber Security\\
Singapore University of Technology and Design\\
Singapore\\
Email: adepu\_sridhar@mymail.sutd.edu.sg, aditya\_mathur@sutd.edu.sg}
}

\maketitle

\begin{abstract}
 A  hackfest named  {\bf S}WaT {\bf S}ecurity {\bf S}howdown (\hackfest)  has been organized  consecutively for two years.  \hackfest  has enabled researchers and practitioners to assess the effectiveness of methods and products aimed at detecting cyber attacks launched in real-time on an operational  water treatment plant, namely, Secure Water Treatment (SWaT).  
 In \hackfest independent attack teams design and launch attacks on SWaT while defence teams protect the plant passively and raise alarms upon attack detection. Attack teams are scored according to how successful they are in performing  attacks based on specific intents while the defense teams are scored based on the effectiveness of their methods to detect  the attacks. This paper focuses on  the first two instances of \hackfest and summarizes the benefits of hackfest and the performance of an attack detection mechanism, named Water Defense, that was exposed to attackers during \hackfest.

\end{abstract}


\pagenumbering{arabic}

\section{Introduction}
\label{sec:introduction}

Industrial Control Systems\,(ICS)\,\cite{stoufferFalcoScarfone} considered in this work are complex interconnected systems deployed to control and monitor, among others, critical infrastructures such as water treatment and electric power systems\,\cite{icsCERTAdvisory}. The increase in successful cyber attacks on public infrastructure\,\cite{germanSteelMill,ukraineBlackout,weinbergerStuxnet}, and other mostly unsuccessful attempts\,\cite{ics-cert}, has raised the importance of deploying cyber defense mechanisms in ICS. 
Attackers are often bypassing the defense mechanisms (prevention and detection) by exploiting software and hardware vulnerabilities or through social engineering. Therefore, it becomes important to look for ways of detecting process anomalies in an ICS caused by an attacker who has gained unauthorized entry. Water Defense\,\cite{patentSridharMathur}, referred to in this paper as \waterdefense,  is one such anomaly detection mechanism. 

The technology underlying \waterdefense has been described and experimentally evaluated by the authors\,\cite{adepuMathurAsiaCCS2016,adepuMathurIFIPSEC2016}.  
This paper describes an independent supplementary  assessment of \waterdefense based on attacks launched by teams of attackers. The assessment was carried out over two consecutive years in an event labeled  ``hackfest (\hackfest)"\,\cite{antonioliGhaeiniAdepuOchoaTippenhauer,s3-16,s3-17}. The event was organized by a team of faculty, students, and staff at the Singapore University of Technology and Design. Several attack and defense teams\,\cite{antonioliGhaeiniAdepuOchoaTippenhauer,s3-16,s3-17} participated in \hackfestend. The attack teams designed and launched attacks on SWaT--the system used for assessment of \waterdefense by the authors. This paper is focused on analysis of data used for assessing the performance of \waterdefense in detecting process anomalies resulting from the attacks. Details of \hackfest are in Section~\ref{sec:eventDetails}. 

The following two research questions were the focus of \hackfest with respect to \waterdefense.  
\skipnoindent{\bf RQ1:} How do attackers compromise the security of an ICS? 
\skipnoindent{\bf RQ2:} How effective is \waterdefense in detecting attacks launched by  independent attack teams? 

\skipnoindent{\it Contributions:} This paper\,(a)~summarizes the performance of the \waterdefense mechanism during two consecutive  \hackfest events  and (b)~reports  on observations and lessons learned from the experience; all attacks launched during \hackfest 
are  reported.  Information presented in this work will likely be valuable to  researchers attempting to assess the performance of process anomaly detection  methods other than \waterdefense. A complete list of the invariants (see Section~\ref{sec:overview}) used in \waterdefense, and the \hackfest dataset, are also available for download\,\cite{itrustDataset}.
 
\skipnoindent{\it Organization:} The remainder of this paper is organized as follows. Section\,\ref{sec:background} contains material to aid in understanding the remainder of this paper. An overview of the \waterdefense is presented in Section\,\ref{sec:overview}. Section\,\ref{sec:eventDetails} contains a description of  \hackfestend.  Preparation for the event is  described in Section\,\ref{sec:preparation}. Attacks performed by attacker teams during both \hackfest events are in Section\,\ref{sec:attackdetails}. Results from the events are presented in  Section\,\ref{sec:s3results}. Section\,\ref{sec:discussion} returns to the research questions mentioned above and answers them with respect to the data generated during \hackfestend. Section\,\ref{sec:relatedwork} is a summary of the work. Conclusions based on \hackfest are summarized in Section\,\ref{sec:conclusion}.


\section{Preliminaries and Background}
\label{sec:background}

This section is a  brief introduction to Industrial Control Systems (ICS)\,\cite{stoufferFalcoScarfone} in the context of SWaT. An example illustrates the nature of a cyber attack on SWaT and its potential impact on system response.

\subsection{Industrial Control Systems}

\begin{figure}[ht]
\centering
\includegraphics[width=\linewidth]{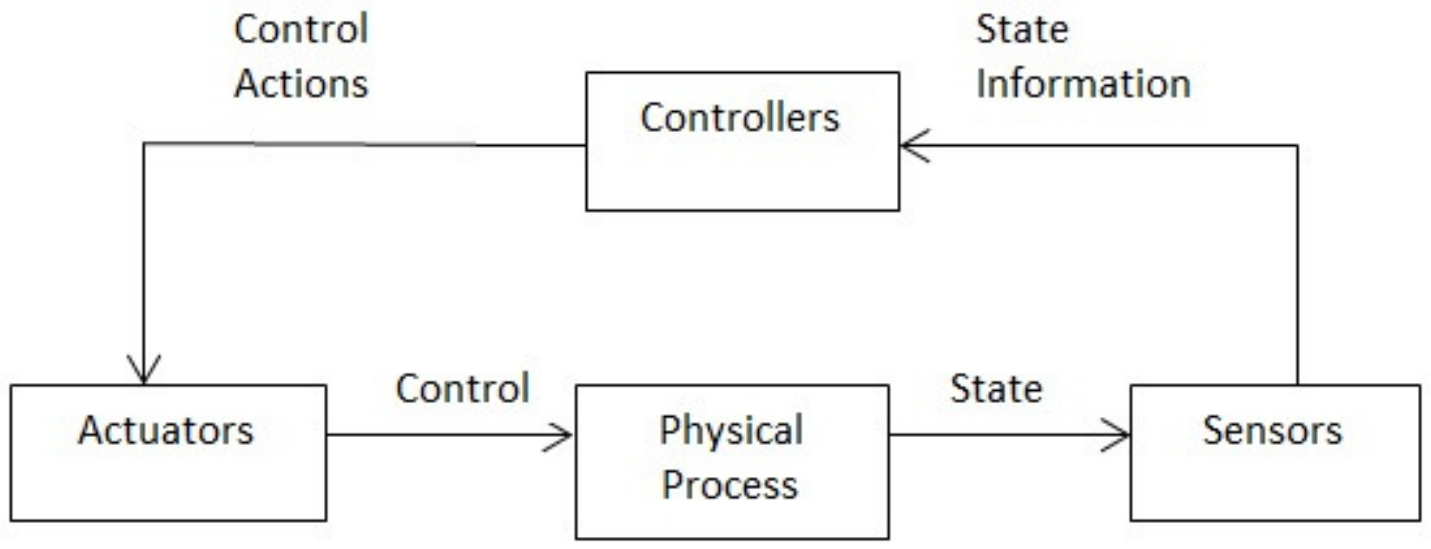}
\caption{\small  High level view of an ICS.\normalsize}
\label{fig:stateTransformer}
\end{figure}

 An ICS\,\cite{stoufferFalcoScarfone} consists of physical,  control, and network  devices (Figure~\ref{fig:stateTransformer}). Physical devices include  sensors and actuators. Control devices include  Programmable Logic Controllers (PLCs), Supervisory Control and Data Acquisition (SCADA) workstations,  and Human-Machine-Interface (HMI) devices. Network devices include network switches and  access points. The PLCs, SCADA, and HMI  monitor and control the physical process. Communication channels in the network act as a bridge between the physical process and the control devices. Communication channels communicate the state of physical process to controllers and control signals to the actuators. PLCs receive data from sensors, compute control actions, and apply these actions to specific devices. The PLCs in an ICS can be viewed collectively as a distributed control system that transforms the state of the process through the use of sensors and actuators.

\begin{figure*}[tbh]
\centering
\includegraphics[width=7in]{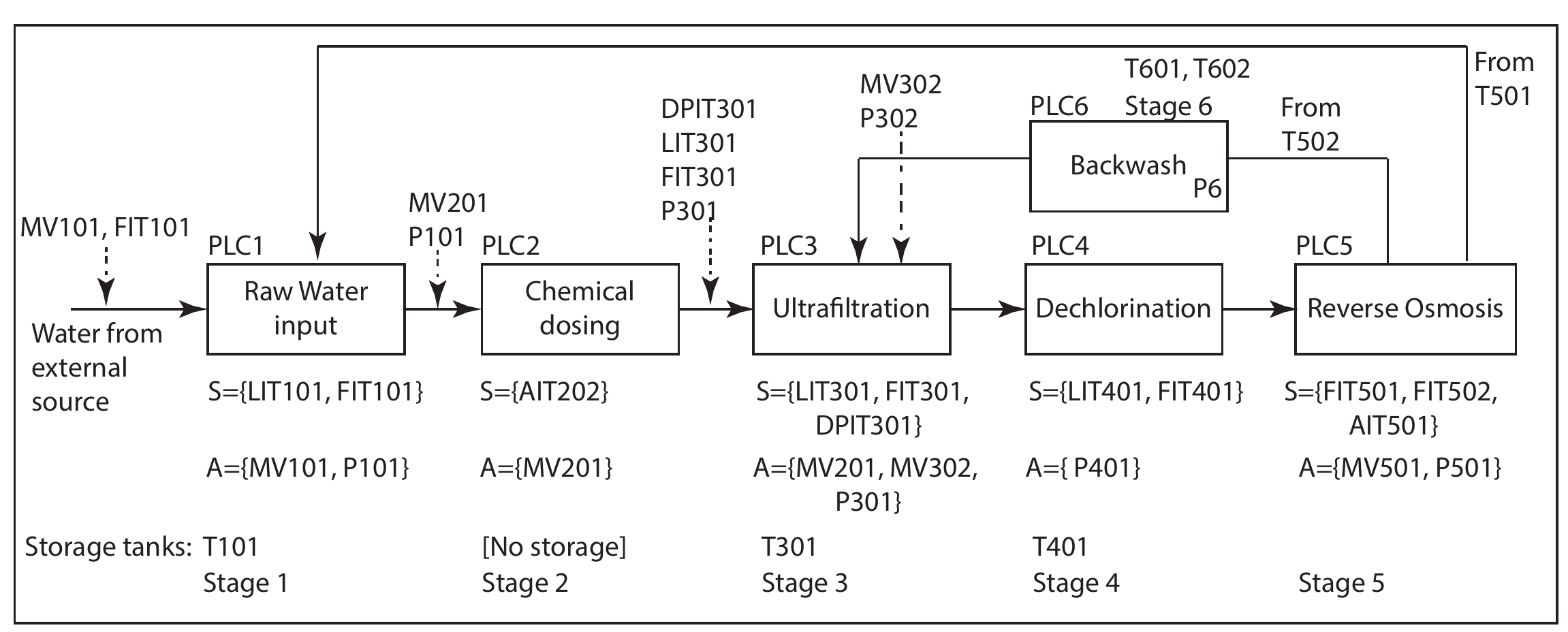}
\caption{Six stages in SWaT with corresponding PLCs, sensors, and actuators; Five water storage tanks as shown are labeled Txxx. Water level in each tank is measured by the corresponding level indicator labeled as LITxxx. AITxxx, FITxxx, and DPITxxx measure, respectively, chemical properties of water, flow rate in a pipe, and differential pressure across the ultrafiltration unit. Pxxx denote pumps at various stages. }
\label{fig:swat-stages}
\end{figure*}

\subsection{SWaT: Architecture and components}

\noindent SWaT\,\cite{mathurTippenhauer,swat} is a testbed for water treatment. It is used to investigate the response to cyber attacks and experiment with novel designs of defense mechanisms such as the ones described in\,\cite{adepuMathurHASE2016}. The architecture and components in SWaT are described next.
\skipnoindent {\em Stages in SWaT}: As shown in Figure\,\ref{fig:swat-stages}, SWaT consists of six stages labeled Stage~1  through Stage~6. Each stage is controlled by its own set of PLCs.  Stage~1 controls the inflow of water to be treated by opening or closing a valve that connects the inlet pipe to the raw water tank (T101). Water from the raw water tank is pumped via a chemical dosing (Stage~2) station to another Ultrafiltration (UF) feed water tank (T301)  in Stage~3. In Stage~3 a UF feed pump sends water, via the UF unit, to a Reverse Osmosis (RO) feed water tank (T401) in Stage~4. In Stage~4 an RO feed pump sends water through an ultraviolet dechlorination unit controlled by a PLC. Differential pressure sensors in Stage~3 measure the pressure drop across the UF unit. A backwash cycle is initiated when the pressure drop exceeds 0.4~bar indicating that the membranes need immediate cleaning. Stage~5 contains a PLC to control the Reverse Osmosis (RO) unit that further filters the water using a 2-stage RO process. The output of the RO unit enters storage tanks T601 and T602 in Stage~6.  Tank T601 contains the {\em reject} from RO and is used to clean the UF unit using a backwash process. Water in tank T602 contains the {\em permeate} which is recycled into tank T101 in Stage~1. 

\begin{figure}[ht]
\centering
\includegraphics[width=0.8\linewidth,height=2.3in]{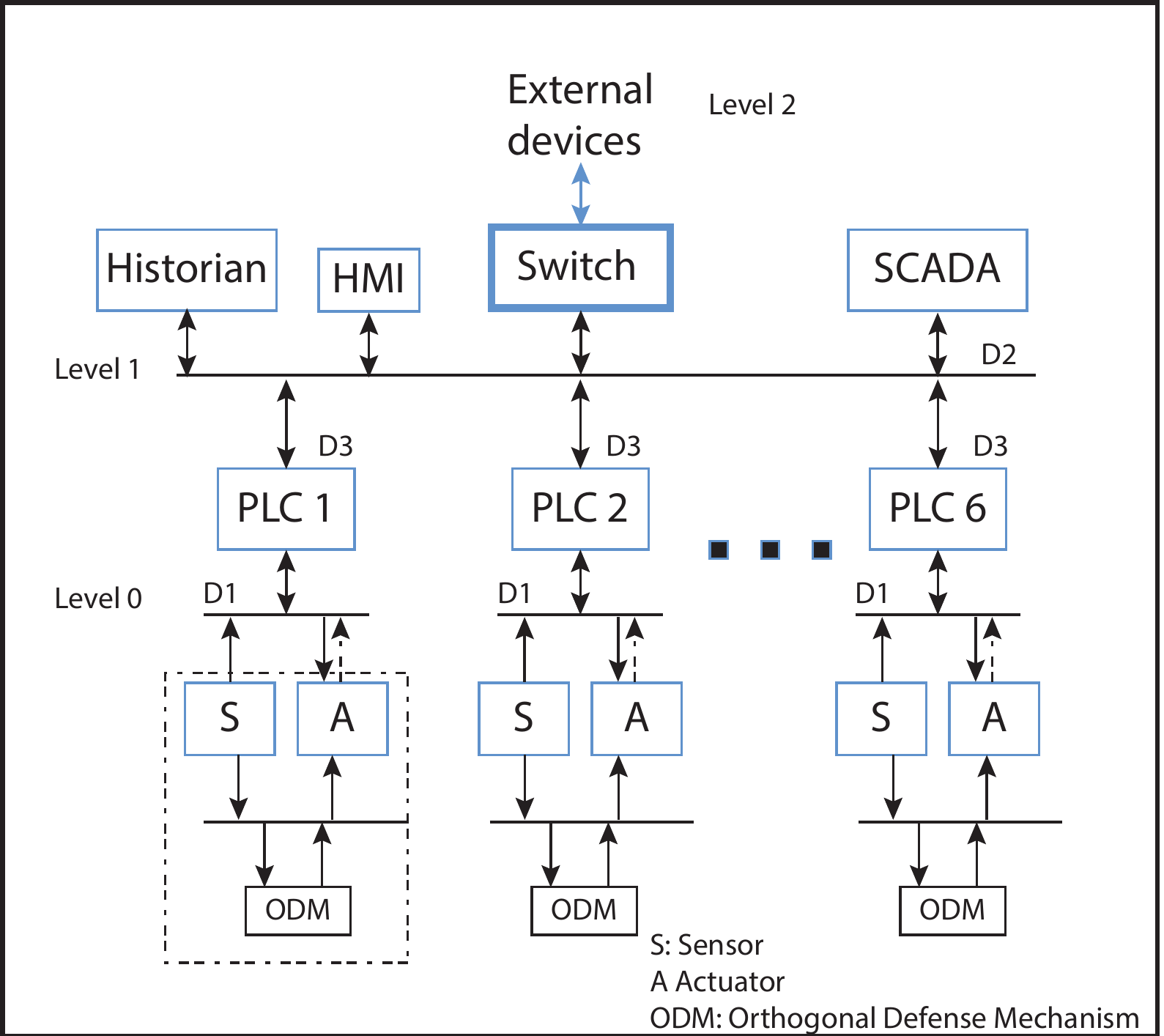} 
\caption{Communications structure of SWaT. D1, D2, and D3 denote sets of defense mechanisms in SWaT; ODM (Orthogonal Defense Mechanism) is independent of these mechanisms. }
\label{fig:commOverview}
\end{figure}

\skipnoindent{\em Sensors and actuators:} SWaT contains 42 sensors and actuators across the six stages. These include sensors that relate to the dynamics of the process such as water level in tanks, flow indicators, and pressure indicators as well as those for  measuring  chemical properties of water including pH, conductivity and hardness. Each PLC has its own set of sensors and actuators connected through a ring network. Thus, when a PLC needs to obtain state information from another PLC, it must request such information via a suitable command; the requested data is sent over level~1 network as shown in Figure~\ref{fig:commOverview}.

\skipnoindent {\em Communications:} Figure~\ref{fig:commOverview} shows the architecture of the communications infrastructure  in SWaT. Each PLC obtains data from sensors associated with the corresponding stage, and controls pumps and valves in its domain. 
Flow of water across various stages is controlled through opening and closing of valves and turning pumps \state{ON} or \state{OFF}.  Level sensors in each tank enable the PLCs to decide when to turn a pump \state{ON} or \state{OFF}. Several other sensors are available to check on the physical and chemical properties of water flowing through the six stages. 
PLCs communicate with each other through a separate network. Communications among sensors, actuators, and PLCs can be via either wired or wireless links controlled manually.  Both wired and wireless networks connect PLCs to the physical process and to the HMI and engineering workstation, i.e. the SCADA workstation.

The control network is  connected to the SCADA workstation\,\cite{swat} through a wired network using a 16-port switch. The PLCs and SCADA workstation  are also connected through  a  wireless network configured using star topology. Switches located at each stage enable the use of either wired or wireless communications. The communications network is layered into two levels. For each PLC, level~0 refers to the communication layer between sensors, actuators and  the PLC. Level~0 network is implemented as a ``device level ring"\,\cite{antonioliGhaeiniAdepuOchoaTippenhauer,UrbanaGiraldoOleCardenas} which  includes a Remote IO (RIO) device. The RIO is connected to the physical sensors and actuators. Monitoring and control information is exchanged between the PLC, sensors, and actuators   across a Distributed Logical Router (DLR).  Level~1 refers to the communication layer  among  PLCs. This layer is  implemented in a  star topology and  includes a SCADA workstation, an HMI, and a Historian.

\skipnoindent {\em Controllers (PLCs):} SWaT is equipped with Allen Bradley ControlLogix PLCs. Therefore, some of the attacks described in the remainder of this paper require consideration of the protocols used by  EtherNet/IP\,\cite{enip} for Allen Bradley PLCs. SCADA software is developed with tools from Rockwell automation\,\cite{rockwellPLCProgramming}.

\skipnoindent {\em Attack detection mechanisms:} As shown in Figure~\ref{fig:commOverview}, SWaT contains four attack detection  mechanisms  D1, D2, D3, and Orthogonal Defense Mechanism (ODM)\,\cite{adepuSiddhantMathur}.  D1 is based on a modified version of the open source Bro intrusion detection tool\,\cite{broIDS}. D2 is a set of three commercially available intrusion detection systems that use a mix of process dynamics and other techniques for anomalous process behavior\,\cite{checkpoint,ics2,kics}. D3 (\waterdefense)  sits inside PLCs and implements a distributed attack detection mechanism that relies exclusively on process dynamics\,\cite{adepuMathurAsiaCCS2016}. The ODM  has direct access to sensors  and analyses the data received  for the existence of process anomaly.

\subsection{An illustrative attack on SWaT}
\label{sec:motiveexample}

Consider Stage~1 of SWaT in Figure~\ref{fig:swat-stages}. This stage has a motorized valve labeled MV101 which, when open, causes water to flow into  tank T101. The  inflow into T101 is measured by flow meter FIT101 and the  water level by a level sensor labeled LIT101. Pump P101 sends water to the next stage.   Flow meter FIT201 measures the outflow of water from Stage~1 to Stage~3. PLC1 receives the LIT101 reading and controls the motorized valve MV101. Similarly, PLC1 receives LIT301 readings from PLC3 and controls pump P101. 

Tanks T101 and T301 have four markers each labeled  Low (\state{L}), Low Low (\state{LL}), High (\state{H}), and High High (\state{HH}). Each marker corresponds to a specific value of water level in  the tank. These markers are used by the corresponding PLCs to control the states of motorized valves and pumps.  Thus, for example,  when the water level in T101 reaches  \state{L}, PLC1 opens  MV101  and closes it when the level reaches \state{H}. When  water level in T301 reaches  \state{L}, PLC1 turns P101 \state{ON}  and turns it \state {OFF} when the level reaches  \state{H}. The following example illustrates the impact of compromising  level sensor LIT101 with the intent of  damaging pump P101.


\skipnoindent{\em  Example:} Consider an attack where the attacker's intention is to underflow T101 and damage P101 by making it run without any incoming water.  The attack is launched on LIT101 with  Stage~1 in the following state: LIT301: 955mm, MV101: \state{Closed}, P101: \state{OFF};  UF is operational and therefore water level in tank T301 is decreasing. Assume now that the attacker sets  LIT101 reading to a constant value of 790mm. In this attack  even though the water level in T101 is changing (decreasing), PLC1 receives a constant value.  After a while when LIT301 reaches \state{L}, pump P101 is turned \state{ON} by PLC1. However, the actual water level in tank T101 is lower than \state{L}, say at \state{LL}. This leads to the outflow from the pump being reduced to less than the intended flow rate. Pump P101 runs dry when there is no water in T101 and  will  eventually get damaged unless a corrective action is taken.

\begin{figure}[ht]
\centering
\includegraphics[width=\linewidth]{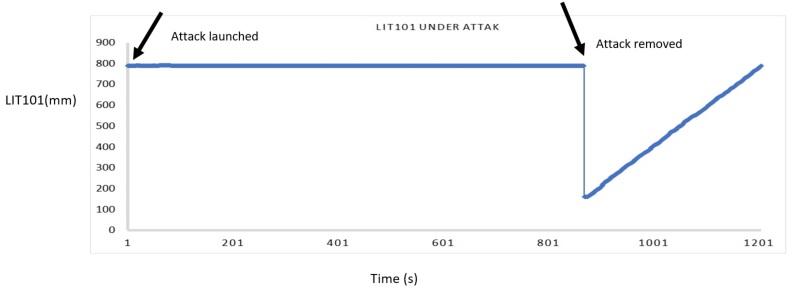}
\caption{\small  Water level in tank T101 when LIT101 is attacked; LIT101 readings are observed by PLC1. \normalsize}
\label{fig:EXP1_LIT101}
\end{figure}

\begin{figure}[ht]
\centering
\includegraphics[width=\linewidth]{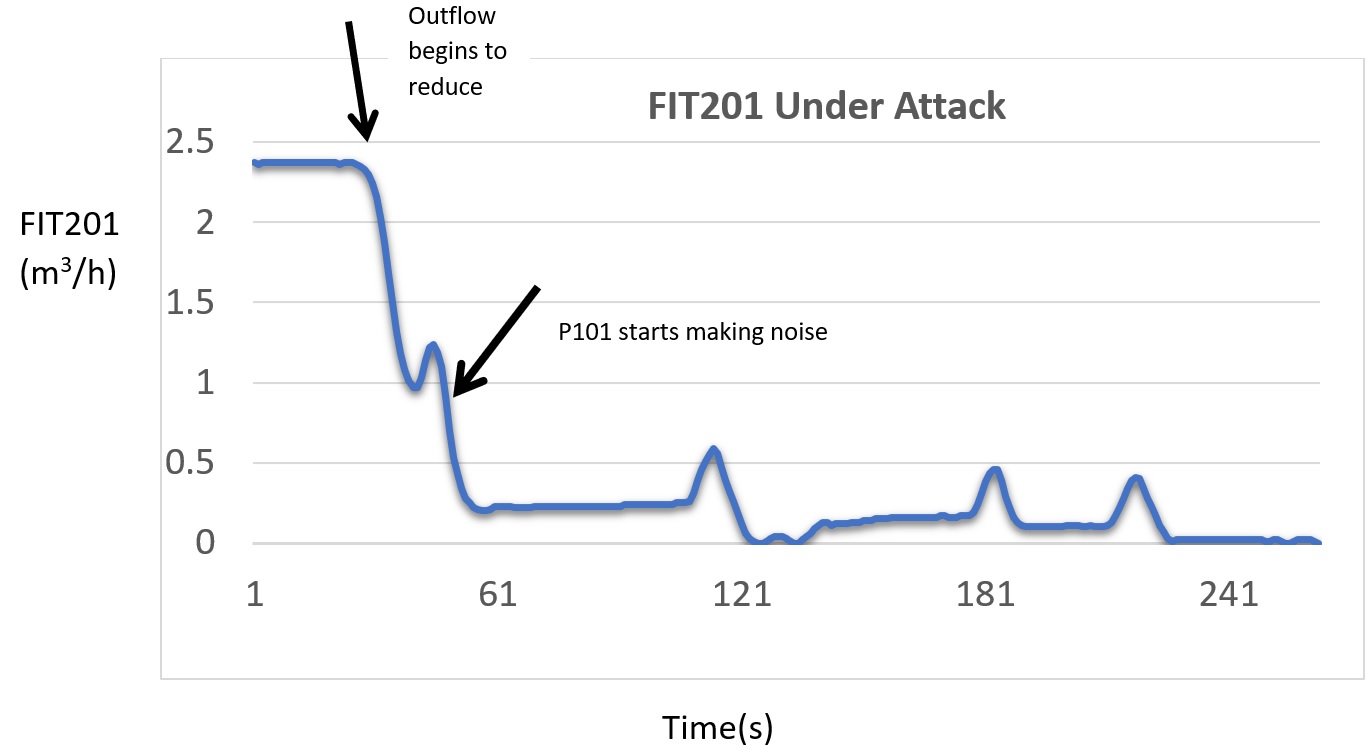}
\caption{\small  Level sensor LIT101 is under attack, The attacker's intention is to underflow T101  tank and damage P101.  The first arrow indicates the outflow reducing time, second arrow indicates the pump noise starting time \normalsize}
\label{fig:EXP1_FIT201}
\end{figure}

Figure~\ref{fig:EXP1_LIT101} shows the water level in tank T101 during the attack.  It can be observed that the outflow increases gradually when the attack is removed. Note that the sudden drop in the value of LIT101 soon after attack removal corresponds to the fact that the PLC begins to receive the correct measurement of water level in T101. When the water level goes down to 150~mm, tank T101 does not have enough water to send to tank T301. Figure~\ref{fig:EXP1_FIT201} shows the change in flow rate during the attack as measured by flow  meter FIT201.  The  two arrows indicate the start of reduction of outflow from T101. At around 10~seconds there is no water flowing from P101 even though the pump is \state{ON}. At this point the pump becomes noisy and the flow rate reduces to zero. If not removed, this attack may lead to pump damage due to overheating. Of course, a mechanical cut off at the pump would avoid such damage.\bend

The above example shows  how an  attacker could potentially damage a pump by changing  the sensor values and actuator states. More  complex  attacks, mentioned in Section~\ref{sec:attackdetails}, can be designed and launched to reduce the chances of being detected.

\section{Overview of \waterdefense}
\label{sec:overview}

\waterdefense is a mechanism to detect process anomalies. A process is considered anomalous when it deviates from its expected behavior. \waterdefense detects such anomalies through the use of invariants.  An invariant\,\cite{adepuMathurIFIPSEC2016} is a condition among physical and/or chemical properties of the process that must hold whenever an ICS is in a given state. At a given time instant, sensor measurements of a suitable set of such properties constitute the observable state of the physical process as known to the ICS.

The invariants serve as checkers of the system state. These are coded and the code placed inside each PLC used for attack detection. The checker code is added to the control code that already exists in each PLC. The PLC executes the code in a cyclic manner. In each cycle, data from the sensors is obtained, control actions computed and applied when necessary, and the invariants
checked against the state variables or otherwise. Distributing the attack detection code among various controllers adds to the scalability of the proposed method. During \hackfest the implementation was located inside the Programmable Logic Controllers (PLCs) as well as embedded in the communication network.

Two types of invariants were considered: state dependent (SD) and state agnostic (SA).  While both types use states to define  relationships  that must hold, the SA invariants are  independent of any state based guard while SD invariants are.  An SD invariant is true when the plant is in a given state; an SA invariant is always true.

\subsection{State-Dependent (SD) invariants}
\label{sec:sdInvariants}

Consider, for example,  the case  when the motorized valve MV101 is \state{Open}. In this case,  the flow rate indicator FIT101 must provide a  non-zero reading to the PLC. This physical fact leads to the following state-dependent invariant: MV101=\state{Open}$\implies$ FIT101$<\delta$, where $\delta$ denotes a threshold indicating flow. Note that an SD invariant may  include conditions from across the various stages of SWaT thus enabling distributed detection of attacks.   Derivation of SD invariants is based on the design of the ICS and is described in\,\cite{adepuMathurIFIPSEC2016}. 

\subsection{State-Agnostic (SA) invariants}

Under normal system operation, an SA invariant must always be true regardless of the system state. One SA invariant was derived for each tank in SWaT to detect attacks that affect the flow of water into and out of a tank. These invariants are based on the flow of water  and water level in a tank, and hence are identical in terms of the mathematical relationship that they capture.  

 As an example of an SA invariant, consider the water level in a tank. 
At time instant $k+1$, the  water level in T101 depends on the level at time $k$ and the inflow  and outflow at instant $k$. This relationship is captured in the following idealized discrete time model of the tank,
\begin{equation}
x(k+1)=x(k)+\alpha(u_i(k)-u_o(k)),\label{invariantIdeal}
\end{equation}
where $u_i(k)$ and $u_o(k)$ denote the inflow and outflow rates at time $k$, and  $\alpha$ is a proportionality constant that converts flow rate to change in level using the tank dimensions. $x(k)$ is the true state of the water level. Let  $y(k)$ denote the  sensor measurement of the water level, $\hat{x}(k)$ an estimate of the level sensor reading, and $\epsilon$  a threshold based on experimentation. Based on Eqn.\,\ref{invariantIdeal},  the statistics obtained experimentally, and converting the  true states to their estimates,  the following invariant is   derived to test whether or not  the tank filling  process is anomalous.

\begin{figure}[h]
\framebox{\vbox{\begin{align}
\frac{\sideset{}{}\sum_{i=1}^{n} |(\hat{x}(i)-y(i))|}{n}&>&\epsilon, & \hskip0.1in\text{under attack,}\label{eq:invariantAttack}\\
&\le&\epsilon, & \hskip0.1in\text{normal.}\label{eq:invariantNormal}
\end{align}
}}
\caption{Invariant to detect anomalous behavior of LIT 101.}
\label{fig:invariant}
\end{figure}

\section{{\bf S}WaT {\bf S}ecurity {\bf S}howdown (\hackfest)}
\label{sec:eventDetails}

This section presents details of the two  \hackfest\,\cite{s3-17}  events including guidelines and selected information on participants. In \hackfest the attackers are challenged to realise concrete goals in  SWaT. Points earned by an attack team are weighted based on the capabilities needed to launch the attack and the number of defence  mechanisms successfully bypassed during the attack. The goal was to meet as many pre-defined challenges as possible within the pre-allocated time.

\skipnoindent{\em Information disclosed to the attack teams}: Technical details on SWaT, such as network architecture, protocols and devices used, are released to the attackers one month prior to their arrival for participation in the event.  Publicly available white papers on mechanisms deployed by the defence teams are  shared with each attack team. 

\skipnoindent{\em Information disclosed to the defenders}: \hackfest organizers worked closely with the defense teams to integrate their defence mechanisms into SWaT. Information about the normal operation of SWaT was disclosed to the defenders to enable them to fine-tune their detection systems and reduce  false alarms as much as they could.

\skipnoindent {\em Attacker profiles}: Attack teams were asked to select from a set of attacker profiles \,\cite{rocchettoTippenhauerESORICS2016}. The following attacker profiles were available:  cyber-criminal, insider,  or a combination of both. An attacker profile is  intended to restrict availability of resources and limit the access rights of the attackers as shown in   Table~\ref{tab:attacker_profiles}.

\subsection{\hackfest-2016}
\label{sec:eventDetails2016}

Attack teams included three from industry and three from academia. Similarly, there were three defense teams from the industry and three from academia. During the live phase, held at the SWaT testbed,  all six\,\cite{s3-16}  defence mechanisms were simultaneously in place.   Each team was given  12~hours for passive reconnaissance and team was assigned  a 3-hour slot during which they were able to launch  attacks. 

\subsection{\hackfest-2017}
\label{sec:eventDetails2017}

Attack teams included one  from industry and four from academia. There were two  defense teams from the industry and two from academia. Each attack team was given two sessions\,\cite{s3-17} of four hours each to conduct reconnaissance on the testbeds. During these sessions, various attacks were prepared and tested with the assistance of the SWaT laboratory engineer. During the actual event, each team was given two hours to demonstrate their attacks that were prepared previously. Attack teams were also given a separate network for Internet up-link and up to three Virtual Machines (VMs) running either Linux or Windows operating system. 

\subsection{Attack targets}

The attack teams were given a list of components and subsystems in SWaT that could serve as the target of their attacks. Table~\ref{tab:attackTargets} lists the targets available to the attack teams. Table~\ref{tab:attackTargets} has two kinds of attacks: physical process attacks and sensor data attacks. In physical process attacks, an attacker's objective is to alter the physical process. In the case of sensor data attacks, an attacker's objective is to alter the sensor or actuator tags during communication or in the  Historian.

\begin{table}
\caption{Resources and access rights for attacker profiles.}
\label{tab:attacker_profiles}
\begin{tabular}{|l| p{2.4in}|}
\hline
{\bf Profile}&{\bf Constraints}\\
\hline
Cyber-criminal&  Limited number of attempts to realize a goal.\\ 
&Physical access not allowed; manual manipulation of the sensors and actuators are not allowed.\\ 
  & Direct connection to PLCs using any software such as Allen Bradley's Studio5000,  not allowed.\\
  \hline
  Insider&Physical access to SWaT allowed; manual manipulation of the sensors and actuators are allowed.\\ 
  &Allowed to alter the network topology\\ 
  & Direct connection to PLCs using any software such as Allen Bradley's Studio5000,  allowed.\\ 
\hline
\end{tabular}
\end{table}

\begin{table}
\caption{targets of attacks in \hackfest.}
\label{tab:attackTargets}
\begin{tabular}{|p{0.8in} | p{2.1in}|}
\hline
{\bf Target}&{\bf Description}\\\hline
\multicolumn{2}{|c|}{\textbf{Physical Process Attacks}} \\ \hline
Valves & Control the motorized valves\\ \hline
Pumps &  Disrupt pump control operations\\ \hline
Pressure &  Alter the pressure in pipes \\ \hline
Tank fill level & Alter water level in a tank\\ \hline

Chemical dosing&Alter chemical dosing\\ \hline

\multicolumn{2}{|c|}{\textbf{Sensor Data Attacks}} \\ \hline 
Historian & Alter data in the Historian \\ \hline
HMI/SCADA &  Alter the sensor, actuator values at HMI or SCADA; DoS Attacks on SCADA, HMI    \\ \hline
PLC		&   Reprogram PLC; DoS attacks on PLCs; Change the commands and values in which the PLC receives and sends \\ \hline
RIO/Display &  Control of the RIO through disconnected analogue Input/Output pin  \\ 
\hline
\end{tabular}

\end{table}

\section{Preparation for \hackfest}
\label{sec:preparation}

To prepare for \hackfest-2016, an earlier version of  \waterdefense  was extended to all six stages of SWaT. This extension required the generation of invariants across all stages, coding of the invariants, and placement of the code inside\,\cite{adepuMathurAsiaCCS2016} the six PLCs. The modified \waterdefense was tested on SWaT by running the plant under various operating conditions.  

Based on lessons learned during \hackfest-2016, several new invariants were generated, coded, and added to the PLCs. For  \hackfest-2017, we decided to use an additional  monitoring system placed outside the PLCs. This system collects  data from the Historian and evaluates the invariants. All invariants were implemented  in a Linux environment using a Piwebclient API to talk to the Historian. This new implementation is referred to as \waterdefenseH. 

The invariants in \waterdefense are coded using ladder logic and structured text, while those in \waterdefenseH in  Python.  Both implementations use the same set of invariants; the difference is in their placement. The  Historian may not get all the data and commands that flow across the PLCs, sensors, and actuators. However, as \waterdefenseH gets its data directly from the Historian, it has access to information flowing  across  SCADA workstation and the Historian. This information may be compromised by an attacker and is not available to the PLC.

\subsection{Scope of \waterdefense}
\label{sec:scope}

\waterdefense is designed to detect process anomalies. Thus, any abnormal behavior in the water treatment process in SWaT ought to be detected by \waterdefense. However, there could be attacks that do not cause the process to deviate from its normal behavior but lead to undesirable consequences. An example of such an attack is one intended to deface the screen on the SCADA workstation or the HMI. Such an attack will not be detected by \waterdefense. Attacks that may cause process anomaly but only after an attack has been removed from the system may also not be detected by \waterdefense. Denial of Service is one such attack.

\subsection{Scope of \waterdefenseH}
\label{sec:scopeH}

\waterdefenseH  and \waterdefense use the same set of invariants. However, the placement of \waterdefenseH could lead to a difference in detection capabilities of the two defense mechanisms. \waterdefenseH gets its data from Historian while \waterdefense directly from the PLC. Data that is not programmed to be logged in the Historian will not be accessible to \waterdefenseH. Thus any anomaly that requires such data will likely not be detected by \waterdefenseH. Similarly, attacks that manipulate data entering the Historian or SCADA may not be visible to \waterdefense.  Thus, while the two invariant-based process anomaly detection mechanisms are identical in the invariants they use, their placement in SWaT  is expected to result in different performance in detecting attacks.

\section{\hackfest Attacks}
\label{sec:attackdetails}
The attacks launched by teams participating in the two \hackfest events are described next.

\subsection{\hackfest-2016 Attacks}

All attacks designed and launched during \hackfest-2016 are enumerated in Table~\ref{tab:2016attacks}. Three attacks selected from Table~\ref{tab:2016attacks} are described next. Details of all attacks are available in\,\cite{antonioliGhaeiniAdepuOchoaTippenhauer}. Of the  18 attacks in Table~\ref{tab:2016attacks},  \ref{dos_attack} and \ref{dos_PLC_SYN_flooding} are cyber criminal attacks and the remaining  are insider attacks.

\begin{table*}[]
\centering
\caption{Attacks launched during \hackfest-2016}
\label{tab:2016attacks}
\begin{tabular}{|c|p{0.8 in}| p{1.5 in}| p{1.7 in}|p{0.5 in}|}
\hline
\textbf{S.NO} &  \textbf{Target} & \textbf{Method} & \textbf{Attack}  & \textbf{Tool} \\ \hline        
\nextattack&    Tank fill level   LIT101  &  Use HMI access   & Close MV101 and Stop P101 and P102& HMI \\ \hline
\nextattack\label{arp_spoofing}& HMI/SCADA  & ARP spoofing &Attack HMI DoS attack   & Ettercap  \\ \hline
\nextattack&PLC  & Manual access &Removed the cable at the ring at level~0& Manual \\ \hline            

\nextattack\label{dos_attack}& HMI/SCADA  & DoS on HMI by dropping all packets between PLC and SCADA/HMI &DoS attack on SCADA, wide DoS attack, took a while to restore SWaT to its normal state & Ettercap      \\ \hline     

\nextattack& Tank fill level LIT101  & Use HMI access &Attack on LIT101& Manual; HMI \\ \hline
\nextattack\label{valve_MV301}&  Valve; MV301 & Use SCADA access &Attack on MV301 manually  open  from the SCAD workstation  & Manual; SCADA     \\ \hline
\nextattack& Pump; P101 &Use SCADA access &Attack  pump; manually open it from the SCADA workstation& Manual; SCADA    \\ \hline
\nextattack& Historian  & DoS attack using CPPPO and loop  &Attack between HMI and PLC& CPPPO    \\ \hline

\nextattack& Valve  MV101& Use SCADA access  & MV101 attacked, using SCADA changed  the valve state from \state{Open} to \state{Closed}& Manual; SCADA    \\ \hline
\nextattack&Pump   P101 & Use SCADA access &LIT301 set point changed & Manual; SCADA\\ \hline
\nextattack& Tank fill level  LIT301 & Using SCADA access &LIT301 set point altered & Manual; SCADA\\ \hline
\nextattack& Chemical dosing  P201 & Control MV101 and AIT503; set points of LIT301 to ensure flow, this triggered chemical dosing & Dosing pump attack on P201& Manual; SCADA\\ \hline

\nextattack & HMI/SCADA,  LIT101 & Functional block, introduce new constant tag, tie that to output tag, could only do zero & LIT101 set to zero from PLC & Studio 5000    \\ \hline

\nextattack\label{chemical_dosing_pump_attack}& Chemical dosing pump  P205 & Use SCADA access  & Manipulation of the chemical dosing pump  (P205)& Manual; SCADA\\ \hline
\nextattack& HMI/SCADA  & DoS on HMI using Level~1 network  & Attack on HMI& Ettercap, \state{Pycomm}         \\ \hline

\nextattack\label{dos_PLC_SYN_flooding}&Historian  & SYN flood ENIP port at PLC1 & DoS to PLC by SYN flooding (attack on HMI) & Ettercap   \\ \hline
\nextattack\label{chemical_p203}& Chemical dosing  pump P203 & HMI-based direct manipulation.  & Attack on P203 while the four dosing pumps are   running& Manual; HMI          \\ \hline
\nextattack& HMI/SCADA  LIT101 & Re-program PLC to fix LIT101 value to an arbitrary value & Attack on LIT101& Studio 5000   \\ \hline
\multicolumn{5}{c}{*\ref{dos_attack},\ref{dos_PLC_SYN_flooding} are cyber criminal attacks in \hackfest-2016}\\ 

\end{tabular}
\end{table*}

\skipnoindent {\em DoS attack on SCADA:} In this attack (attack \ref{dos_attack} in Table~\ref{tab:2016attacks}) the attacker's intention was to deface the SCADA workstation screen and hence prevent the operator from observing plant state.  The cyber-criminal attacker model was used to design this attack. To realize the intention, the attacker launched  an ARP poisoning Man-in-the-Middle attack in two steps. In the first step  all traffic intended for HMI was  redirected to the SCADA workstation.  In the second step, this redirected traffic was dropped and thus no packets were received at the SCADA workstation.  This led to the screen on the workstation becoming completely gray and  no state information was displayed. This attack was not detected by \waterdefense  as it did not lead to any process anomaly. It is an ARP spoofing attack, and  not a traditional DoS attack. As part of the DoS attack, the attacker targeted the PLC and sent millions of packets at a time. This led to the same effect as would be the case when an ARP spoofing attack is performed on SCADA.

\skipnoindent {\em Manipulation of the chemical dosing pump:} Intention of the attacker in this case  (attack \ref{chemical_dosing_pump_attack} in Table~\ref{tab:2016attacks}) was to manipulate  the pH of water entering Stage~3 of SWaT. The insider-attacker model was used in the design of this attack. This attack was executed in two steps. In the first step, PLC\,2 was set to manual mode. Note that in manual mode the plant operator can directly control the actuators, e.g., the dosing pumps in this case. In the second step, the attacker  altered  the chemical dosing process in the Pre-treatment Stage~2 of SWaT by interacting directly with the HMI interface and overriding the commands sent by the PLC. \waterdefense was able to detect this attack because the set-points changed by the attacker were different from those  set in \waterdefense.

\skipnoindent {\em DoS to PLC by SYN flooding:} The intention of the attacker in this case (attack \ref{dos_PLC_SYN_flooding} in Table~\ref{tab:2016attacks}) was to disable the HMI so that an operator is unable to view or control the plant operation. The insider-attacker model was used in the  design of this attack.  In this way the attacker had an access to the administrator account and the associated tools. The attacker performed a SYN flooding attack on Ethernet/IP server of PLC1. 

As a result of this DoS attack, the HMI was  unable to obtain the current state values to display, and would instead display 0 or * characters. \waterdefense was unable to detect this attack   physical process as not affected. During the attack period, PLC was controlling the process as expected. Such attacks, while not altering process behavior, may impede  supervision of the process in  an operational plant.

\subsection{\hackfest-2017 Attacks}
\label{sec:2017attacks}

All  attacks designed and launched during \hackfest-2017 are enumerated in Table~\ref{tab:2017attacks}.  Selected attacks from Table~\ref{tab:2016attacks} are described next. Details of all attacks are available in\,\cite{s3:2017Report}. 
Of the 31 attacks in Table \ref{tab:2017attacks}, 17 can be classified as  cyber criminal attacks and the remaining as  insider attacks (Figure~\ref{tab:attacker_profiles}). All attacks  launched during  \hackfest-2016 and \hackfest-2017 are listed and categorized  in Table~\ref{tab:cybercriminal-attacks}.

\setcounter{attack}{0}
\begin{table*}[]
\centering
\caption{Attacks launched during \hackfest-2017}
\label{tab:2017attacks}
\begin{tabular}{|c|p{1.3 in}|p{1.5 in}|p{2.1 in}|p{1.4 in}|p{0.5 in}|}
\hline
\textbf{S.No} & \textbf{Target}  & \textbf{Method}  & \textbf{Attack}  & \textbf{Tool}    \\ \hline
\nextattack& HMI/SCADA, LIT401& HMI simulation insider attack & Change the value of LIT401 in the HMI & Manual; HMI  \\ \hline
\nextattack& Historian& ARP and drop& Change the value stored at the Historian & Ettercap   \\ \hline
\nextattack& Valve   MV201& Reprogram PLC & Change the status of the MV201 & Studio 5000   \\ \hline
\nextattack& Tank fill level LIT301, 420 to 320 &  Manual & Lower the water tank level from 820mm to 420mm without raising any alarm; LIT301 decreased till 320mm     & Manual; HMI    \\ \hline
\nextattack\label{attack:pump}& Pump P101 & Manual mode of pump  & Alternate the state [On:Off] of the pump P101 & Manual; HMI  \\ \hline
\nextattack& Chemical dosing   P205 & Manually dosing chemical pump  & Change the chemical dosage of sodium hypochlorite (NaOCl) in P2 & Manual; SCADA    \\ \hline
\nextattack& PLC& Disconnect cable & Disrupt sensor values from remote input/output (RIO) to the PLC & Manual               \\ \hline
\nextattack& RI/O Display& Disconnect IO PIN manual& Disrupt the sensor reading send to PLC through Remote I/O (RIO) & Manual \\ \hline
\nextattack& Chemical dosing P404 & MiTM, Python script to control& Increase chemical dosage in pre-treatment & Python script\\ \hline
\nextattack & LIT101 (476mm to  540mm )    & Reprogram PLC  & Falsify water level display at SCADA  & Studio 5000  \\ \hline
\nextattack& Pump P101& HMI simulation insider attack  & Alternate the state [On:Off] of the pump P101  & Manual; HMI    \\ \hline
\nextattack& HMI/SCADA AIT 504& ARP+rewriting. & Increase AIT504 & Ettercap    \\ \hline
\nextattack& PLC LIT401 & Reprogram PLC & Falsify water level display at SCADA   & Studio 5000  \\ \hline
\nextattack& RIO/Display& Disconnect specific IO PIN based on manual  & Disrupt the sensor reading send to PLC through remote I/O (RIO)  & Manual    \\ \hline
\nextattack\label{chemical_dosing_p403}& Chemical dosing pump   P403, AIT501  & Based on captured traffic between HMI and PLC4 & Change chemical dosing function & VNC, Python script, \state{Pycomm}, \state{Wireshark}  \\ \hline
\nextattack\label{PLC_MITM}& PLC, LT101 from 742mm  to 500mm  & Level~0 MITM  & Change the commands and values that the PLC receives and sends  & Aircrack, Airodump, Aireplay, \state{Netfilterqueue}, \state{Scapy}    \\ \hline
\nextattack\label{Historian_aircrack}& Historian, LT101 tag  & Aircrack WiFi; ARP spoofing, Ettercap & Compromise historian data  & Ettercap, Aircrack   \\ \hline
\nextattack\label{pressure_smb}& Pressure sensor  DPIT301/30, MV301-4& SMB to EW, get project files, run FT & Disrupt valves operation of Ultrafiltration and Backwash (P3) &  SMB  \\ \hline
\nextattack\label{level_metaexploit}& MV201, LT101&metasploit+vnc & Change the water level of the tank; LIT101  & Metasploit+vnc   \\ \hline
\nextattack\label{pump_rougue}& Pump P501& Rogue AP disassociated; Telnet with default credentials to turn off original AP. \state{Scapy} rewrite. & Disrupt pump control operation  & KisMAC, Password cracking tool, 3vilTwinAttacker, Telnet, \state{Scapy}  \\ \hline
\nextattack& PLC, LIT101& Reprogram PLC& Change level indicator value & Studio 5000 \\ \hline
\nextattack& Pump P101, LIT301& Using back-door connection & Establish back-door connection & Mimikatz, malicious VBA Macro, SOCKS proxy \\ \hline
\nextattack& HMI/SCADA P201& \state{Netfilterqueue}, \state{Scapy} & Change the display value of the HMI & \state{Netfilterqueue}, \state{Scapy}   \\ \hline
\nextattack& Historian LIT101&  & Overwrote specific data stored at the Historian & Microsoft PsExec, ipconfig   \\ \hline
\nextattack& RIO/Display& Manual & Control of the RIO through disconnected Analogue Input/Output pin & Manual \\ \hline
\nextattack& Valve MV201& Manual  & Permanently closed the motorised valve regardless of commands issued &  Manual; SCADA   \\ \hline
\nextattack& RIO/Display AIT202/203& Manual& Change the pH value shown at HMI & Manual; HMI   \\ \hline
\nextattack& PLC, MV201, P101 & ARP poisoning, MiTM & Increase the pressure at P1  & Ettercap   \\ \hline
\nextattack& Tank  fill level  LIT101& Lower& Falsify the water level reading of the tank displayed at SCADA & \state{Pycomm}    \\ \hline
\nextattack& Chemical dosing,  PLC2& Use Studio 5000 & Change the level of the chemical used for dosing & Studio 5000  \\ \hline
 \nextattack& Pressure, MV302, P301/2& Using \state{Pycomm} script & Change the pump state sent to the PLC& \state{Pycomm}   \\ \hline
\end{tabular}
\end{table*}

\begin{table}
\centering
\caption{Cyber criminal attacks in \hackfest}
\label{tab:cybercriminal-attacks}
\begin{tabular}{p{0.3 in}|p{1 in}|p{1 in}|} \cline{2-3}

&\mc{1}{c}{\bf Cyber Criminal Attacks} & {\bf Insider Attacks} \\ \hline

\mc{1}{|l|}{\textbf{\hackfest-2016}} & 4, 16  & 1, 2, 3, 5, 6, 7, 8, 9, 10, 11, 12, 13, 14, 15, 17, 18     \\ \hline

\mc{1}{|l|}{\textbf{\hackfest-2017}}& 2, 9, 10, 12, 13, 15, 16, 17, 18, 19, 20, 21, 22, 24, 28, 29, 30    & 1, 3, 4, 5, 6, 7, 8, 11, 14, 23, 25, 26, 27, 31    \\ \hline        
\end{tabular}
\end{table}







\skipnoindent{\em  Control of the chemical dosing system through a Python script (\state{Pycomm}):} The objective of this attack (attack\,\ref{chemical_dosing_p403} in Table\,\ref{tab:2017attacks}) was to change  chemical dosing at the end of the de-chlorination system (Stage~4). First, the  attackers  compromised  Virtual Network Computing (VNC). Then they used a  Python script (\state{Pycomm}) and  \state{Wireshark} to gain access to the HMI.  After gaining access to the HMI through the compromised VNC, the cybercriminal attacker used \state{Wireshark} to capture the  packets flowing between the HMI and PLC4. The controller tags were retrieved by an analysis of the packets.   The attackers changed the data associated with these tags to control the chemical dosing function using the \state{Pycomm} framework.

\skipnoindent {\em Control of  PLC through the Bridged Man-in-the-Middle (MiTM) at Level\,0:} the objective of this attack was (attack\,\ref{PLC_MITM} in Table\,\ref{tab:2017attacks}) to change the commands and values that PLC1 receives and sends. First, the attackers configured a bridge between the RIO and PLC1 using \state{Netfilterqueue} and  \state{Scapy}. The attack was launched at two network levels.  An analysis on the network traffic revealed the packets that the attackers should edit. As the target of this attack was the water level in T101, the attackers set it to a constant value to hide from PLC1 the rise in water level in T101. Before a packet was forwarded, \state{Netfilterqueue} rerouted it into a queue which can be read and modified by the Python script. To prevent all packets from entering  the queue, in order not to disrupt other processes, \state{iptables} was used to identify the targeted packets entering the queue. Using \state{Scapy} and a custom dissector, the attacker edited the payload of the targeted packet which was then forwarded to its original destination.

\skipnoindent{\em  Control of  Historian through the Aircrack WiFi:} The objective of this attack was (attack\,\ref{Historian_aircrack} in Table\,\ref{tab:2017attacks}) to compromise the data stored in the Historian. Attackers performed crack WiFi password, ARP poisoning, and MiTM payload manipulation using \state{Aircrack} and \state{Ettercap}. As PLC1 was operating in the  wireless mode, the cybercriminal attacker used \state{Aircrack} to obtain the password for connecting to the ICS Access Point (AP). ARP poisoning was executed to reroute traffic between PLC1 and the Historian through the attacker's rogue terminal. The attackers then used an \state{Ettercap} filter to manipulate the network packets. The attackers changed the tag corresponding to LIT101 to an arbitrary value before releasing the packets to the Historian. 

\skipnoindent{\em  Control of  pressure through the Server Message Block (SMB):} The objective of this attacks was (attack\,\ref{pressure_smb} in Table\,\ref{tab:2017attacks}) to disrupt the state of four motorized valves in Stage~3 to affect the differential pressure in UF.  Vulnerability CVE-2008-2160\footnote{https://www.cvedetails.com/cve/CVE-2008-2160/} in Factory Talk software from Rockwell,  and in Microsoft's Server Message Block (SMB),  was used by the attackers to obtain files from the HMI. As the HMI was running Windows CE, it has a vulnerability that allows an attacker's terminal to execute arbitrary code on the HMI. Thus the attackers were  able to retrieve the files to create a copy of the workstation. From the copied workstation, the attackers manually changed the state of the valves in Stage~3 such that the  differential pressure  across the UF unit as measured by DPIT301, became dangerously high. The attackers closed  valves MV301, MV302, and MV303, and opened MV304.  

\skipnoindent{\em Control of water level in the tank through the \state{Metasploit VNC Scanner}:}  Objective of this attack was (attack\,\ref{level_metaexploit} in Table\,\ref{tab:2017attacks}) to change the water level in  tank T101. The attackers used \state{Metasploit VNC} authentication \state{None} scanner to obtain  access to the VNC server without password protection and to check for nodes running a VNC Server. Once the scanner detected the VNC Server running without any authentication, the attackers penetrated into the server through a VNC Client connection. As the VNC Server was hosting the HMI which controlled the ICS, the attackers changed the simulation tag associated with water level in T101. 

\skipnoindent{\em  Control of a pump through a rogue router:} The objective of this attack (attack\,\ref{pump_rougue} in Table\,\ref{tab:2017attacks}) was to disrupt the control of pump P501.  The attackers used Evil twin (rogue access point) method using \state{KisMAC}, a password cracking tool, \state{3vilTwinAttacker}, \state{Telnet}, and \state{Scapy}. The  attackers used \state{KisMAC} to scan for wireless networks in the ICS. Once the targeted wireless network was identified, the attackers used dictionary attack to crack the password.  After the password was cracked, the attackers created a rogue wireless router with a similar SSID and configuration. They then  sent a de-authentication packet to disassociate  PLC5 and the original router. The attackers used \state{Telnet} to log into the original router and shut it down. \state{Scapy} was then used to modify the packets to turn the pump on.

\section{Results}
\label{sec:s3results}

Tables~\ref{tab:2016results} and ~\ref{tab:2017results} summarize the response of \waterdefense and \waterdefenseH to the attacks launched during the two \hackfest events.  Recall that both \waterdefense and \waterdefenseH contain exactly the same set of invariants. In \waterdefense the invariants are coded and placed inside the PLCs whereas in \waterdefenseH the invariants are coded and placed at the Historian. \waterdefenseH did not exist during \hackfest-2016 and hence the response of \waterdefenseH is available only for attacks launched during \hackfest-2017.

\subsection{\hackfest-2016 results}

\setcounter{attack}{0}
We note from Table~\ref{tab:2016results} that 10 out of 18 attacks were detected immediately while the remaining eight attacks were not detected. Six  of the eight undetected attacks did not lead to process anomaly during the observation period and hence did not violate any invariant.  This outcome is expected as the invariants in \waterdefense are designed to detect process anomaly.

Consider  attack \ref{arp_spoofing}, ARP spoofing, in Table~\ref{tab:2016attacks}. This is a DoS attack on HMI. It leads to defacing the screen on the HMI, or displaying incorrect information,  thereby preventing an operator from knowing the actual plant state.  However, the attack does not cause process anomaly and hence is not detected as it  does not violate any invariant. Similar logic can be used to explain why the  other attacks in Table~\ref{tab:2016results} are not detected. 

It is important to note that a DoS attack, when given enough time to evolve and be launched at an appropriate state of the plant, may impact physical process behavior. In such a case   one or more invariants may detect the attack. One such attack is \ref{dos_PLC_SYN_flooding} in Table~\ref{tab:2016results}. This attack prevented the  Historian from receiving data from PLC1. However, if this attack was left active for a longer period, it would prevent PLC1 from sending appropriate commands to the actuators, e.g., to MV101 or P101. In turn this would have led to process anomaly. Not enough data is available to conclude with certainty whether or not this attack would be detected by \waterdefense if active for  sufficient time.

Two single point\,\cite{adepuAttackerModel} attacks were not detected by \waterdefense. In one attack (attack\,\ref{valve_MV301} in Table\,\ref{tab:2016attacks})  the adversary altered the status of valve MV301. Under normal circumstances this valve is opened during the backwash process. However, the attacker opened it when there was no backwash. Hence the attack did not affect the physical process except in changing the valve status. No invariant was violated due to this attack because the backwash process, i.e., Stage~6, is not included in this case study. The second single point attack (attack\,\ref{chemical_p203} in Table\,\ref{tab:2016attacks})  was performed on chemical dosing pump P203 while the other pump P204 was running. Note that under normal circumstances only one of these two pumps is supposed to be running while the other remains as a backup. Subsequently the attacker shut down pump P204. This attack was not detected because there were no invariants that related to the chemical properties of water.



Although the overall performance of \waterdefense was below 100\%, it did detect all attacks within its scope except two (attacks 6 and 17 in Table III) as mentioned earlier.

\subsection{\hackfest-2017 results}

Table\,\ref{tab:2016results} indicates that 21 out of 31 attacks were detected by \waterdefense while 24 out of 31 attacks were detected by \waterdefenseH. 
Considering only the attacks within its scope, as mentioned in Section~\ref{sec:scope}, \waterdefense detected 21 out of 28 attacks (75\%). Similarly, \waterdefenseH detected 24 out of 31 attacks (77.41\%) within its scope mentioned in Section~\ref{sec:scopeH}. Three attacks on the Historian are not in the scope of \waterdefense. All attack targets related to RIO/Display (in Table\,\ref{tab:attackTargets} and in Table\,\ref{tab:2017attacks}) are not detected by both  \waterdefense and \waterdefenseH. This is because registers inside a PLC save  the previous values received from the sensors, and the PLC continues to execute the control code. The  invariants also use the same values stored in the PLC registers and hence do not raise an alert.  

In general, PLCs send to the Historian, via the SCADA workstation, the data received from the sensors. When a PLC does not have updated values during the attack period, it is obvious that the Historian also receives the same stale values. This is the reason why \waterdefenseH also did not detect attacks related to RIO/Display. Note that the  RIO/Display attacks were launched and remained active only for a  few seconds. During this period the PLC did not update the current sensor values coming through the RIO.  If the same attack is performed for a longer duration, the PLC would update the data received from the sensors. Doing so would likely lead to \waterdefense and  \waterdefenseH detecting  the RIO attacks.

Attacks launched on the Historian were detected by \waterdefenseH but not by \waterdefense. This variance is due to the fact that data in these attacks is manipulated at the Historian. Thus, invariants in a PLC do not have access to the manipulated data and hence the invariants in \waterdefense do not raise any alert.  All attacks targeting a PLC are detected by \waterdefense and  \waterdefenseH.


\begin{table}
\centering
\caption{Performance of \waterdefense and \waterdefenseH}
\label{tab:2016results}
\begin{tabular}{p{0.4in}|p{0.5in}|p{0.7in}|p{0.8in}|}
\cline{2-4} 
&\mc{1}{p{0.5 in}}{\bf\hackfestend-2016}&\mc{2}{|c|}{\bf\hackfest-2017}\\\cline{2-4}
 & \mc{1}{p{0.5in}|}{\bf \waterdefense}&  \mc{1}{p{0.7in}|} {\bf \waterdefense} & \mc{1}{p{0.8in}|}{\bf \waterdefenseH} \\ \hline
\mc{1}{|p{0.5in}}{\textbf{Detected}} &\mc{1}{|p{0.5in}|}{ 1, 5, 7, 9, 10, 11, 12, 13, 14, 18} & \mc{1}{p{0.7in}|}{ 3, 4, 7, 9, 10, 11, 12, 13, 15, 16, 18, 19, 20, 21, 22, 23, 26, 28, 29, 30, 31}  & \mc{1}{p{0.8in}|}{ 2, 3, 4, 7, 9, 10, 11, 12, 13, 15, 16, 17, 18, 19, 20, 21, 22, 23, 24, 26, 28, 29, 30, 31 }  \\ \hline

\mc{1}{|l|}{\textbf{Not detected}}& 2, 3, 4, 6, 8, 15, 16, 17 & 1, 2, 5, 6, 8, 14, 17, 24, 25, 27   &  1, 5, 6, 8, 14, 25, 27  \\ \hline        
\end{tabular}
\end{table}

\skipnoindent{\em \waterdefense: Detection of physical process attacks:}
All attacks on  valves, pressure sensor,  and level sensors were detected. Three out of four attacks on the chemical dosing process pumps were detected. An example of a detected attack is when the attackers took control of  pump P301 (attack\,\ref{pump_rougue} in Table\,\ref{tab:2017attacks}) through a Python script (\state{Pycomm}) to raise the pressure in the UF unit, measured by  sensor DPIT301, to a dangerous level. \waterdefense immediately raised an alarm. This invariant  ensured that  pump P301 must be \state{OFF}   when the pressure at DPIT301 was above a threshold. During the attack the invariant  was violated as the pump was not turned off while DPIT301 indicated readings that were above the  threshold. Consequently an alarm was raised immediately. In certain cases, multiple alarms were raised due to the violation of one or more invariants. For example, when level sensor LIT101 was compromised, the invariants corresponding to this sensor were violated and raised alarms. 


\skipnoindent{\em \waterdefense: Detection of sensor data attack:}
\waterdefense detected attacks on HMI/SCADA and PLC values because these attacks directly compromised the physical processes. These attacks either compromised chemical dosing, water tank levels, or pump status through hacking of the HMI/SCADA or PLC. Hence, the robustness of \waterdefense in detecting unusual physical process behavior was found effective in these attacks. On the other hand, \waterdefense was unable to detect insider attacks that  pulled out RIO cables. This is because \waterdefense  triggers an alarm only when the invariants are violated. Under normal circumstance, for a period of time, a PLC continues to execute its control code, and any invariant code based on the last known state and/or values. Thus the invariants located inside the PLCs are unable to observe this anomalous behavior.

\skipnoindent{\em\waterdefenseH: Detection of physical process attacks:}
\waterdefenseH  detected 14 out of 16 physical process attacks.   


\skipnoindent{\em \waterdefenseH:	Detection of sensor data attacks:}
\waterdefenseH detected the attacks on HMI/SCADA and PLC values because these attacks directly compromised the physical processes, albeit with a slightly lower detection rate when compared with the rate of detecting physical process attacks. As with  \waterdefense, \waterdefenseH did not detect any attack launched against the Remote I/O by pulling the cables that connect it to the corresponding PLC. 
\waterdefenseH fared better in the detection of attacks against the Historian as it was directly accessing data on the Historian server. 
\skipnoindent If the Historian itself, or data that is input to the Historian is compromised,  WDH takes the decision based on the input  it receives. A clever and  powerful attacker can attack the physical process and modify values entering the  Historian and thus deceive WDH.  In general, such a situation may arise in all behavioral intrusion detection systems where the detector takes the decision based on incorrect input data.

Indeed,  data that appears to be ``legitimate" could lead the WDH into believing that there is nothing wrong with the physical process though there actually is. However, doing so requires the attacker to continuously manipulate a large number of state variables. For example, consider an attack where the attacker turns a pump, say P101, \state{ON} when it should be \state{OFF} and (continually) sends the state of the pump as \state{OFF} to the Historian and the corresponding PLC. If the pump is \state{OFF} then the level of the source and destination tanks must be, respectively, decreasing and increasing at rates determined by the pump characteristics. Creating ``legitimate-looking" data thus requires an attacker to manipulate several state variables as explained next. (a)~Two state variables that correspond to tank levels.  Two sensors (in SWaT) measure these state variables (see Figure~\ref{fig:swat-stages}). Thus, the attacker must have access to these level sensors. (b)~If pump P101 is actually \state{ON} while the Historian receives its state as \state{OFF}, then FIT201 must show no flow. Thus the attacker will also need to manipulate FIT201 to avoid detection. This argument can be carried forward to subsequent stages to show that many sensors will need to be manipulated by an attacker to ``hide" a simple attack such as ``change the state of a pump." In summary, yes, incorrect data at the Historian could prevent detection though doing so would be a significant challenge for the attacker due  primarily to the distributed nature of the invariants.

\begin{table}[]
\centering
\caption{ Results from \hackfest 2017}
\label{tab:2017results}
\begin{tabular}{|p{1.2in}|p{0.4 in}|p{0.3 in}|p{0.3 in}|}
\hline
\textbf{Target of Attack}& \textbf{No.of attacks} & \textbf{\waterdefense} & \textbf{\waterdefenseH} \\ \hline
\multicolumn{4}{|c|}{\textbf{Physical Process Attacks}} \\ \hline
State of motorised valves  & 2 & 100\%& 100\% \\ \hline
State of water pumps& 4& 75\%&75\%   \\ \hline
Pressure in UF&  2& 100\%& 100\%   \\ \hline
Water tank level& 4 & 100\%& 100\%   \\ \hline
Chemical dosing& 4& 75\%& 75\%  \\ \hline

\multicolumn{4}{|c|}{\textbf{Sensor Data Attacks}}\\ \hline
Data in historian & 3& 0\%& 100\% \\ \hline
Data in HMI/SCADA   & 3  & 67\% & 67\%   \\ \hline
Tampering PLC  communications  & 5 & 100\%  & 100\%   \\ \hline
Tampering Remote I/O  & 4  & 0\%  & 0\%   \\ \hline
\textbf{Total Attacks}& \textbf{31}  & 67.74\% & 77.41\%   \\ \hline
\end{tabular}
\end{table}

\section{Discussion}
\label{sec:discussion}

\subsection{Challenges faced}

We faced several challenges during \hackfestend. For example,  after each team's performance, the operator was required to bring  SWaT back to a predefined  normal state. It was necessary to keep SWaT in a normal state before another  team launched attacks. Bringing SWaT to its normal state required: (a)~resetting network communications to ensure that all the communication channels are operating as expected, (b)~the operator to ensure that all physical processes in SWaT are  stable with respect to the control logic, (c)~the operator to bring back SWaT to the normal state of that particular device such as a pump or a motorized valve in the case of  any physical or  manual attacks by the previous team, and (d)~that the Historian and  SCADA servers were reverted to their original state, i.e, the state that existed prior to the launch of attacks.

\subsection{Research questions}

\skipnoindent{\bf RQ1:} How do attackers compromise the security of an ICS? In Section\,\ref{sec:attackdetails} we presented and categorized the attacks based on  attacker profiles. An attacker can launch physical attacks when inside the plant such as manually operating a  motorized valve or tampering with network cabling.  Several  attacks launched by the attack teams had not been launched by the authors in their evaluation of \waterdefense\,\cite{adepuMathurAsiaCCS2016} and \waterdefenseH. Thus, \hackfest raised our confidence in the effectiveness of the attack detection mechanisms  based on invariants derived from plant designs.

\skipnoindent{\bf RQ2:} How effective is \waterdefense in detecting attacks launched by  independent attack teams?  As mentioned earlier, while both \waterdefense and \waterdefenseH were found to detect a number of attacks, they did fail in several cases. Given that the invariants derived are intended to detect process anomalies, it is clear that such mechanisms must be used in conjunction with other attack detection tools such as those in \,\cite{ics2,kics,GhaeiniTippenhauer}.

\subsection{Assessment by the authors and by independent teams}

Table~\ref{tab:performanceComparison} lists the number of attacks launched by the authors in an experimental evaluation performed prior to \hackfest-2016\,\cite{adepuMathurAsiaCCS2016}. Note that the \waterdefense detection rate observed by the authors (89\%) was higher than the combined rate observed during the two \hackfest events (63.26\%). The difference in performance is due to different attack vectors used in the three sets of experiments. \waterdefenseH detection rate observed during \hackfest event is (77.41\%), which is much higher than the \waterdefense detection rate. Some of these attack vectors are explained in Section~\ref{sec:attackdetails} and the remaining may be found in\,\cite{s3:2017Report}.

\begin{table}[h]
\caption{Performance of \waterdefense as evaluated by the authors   against those by participants in \hackfest.}
\label{tab:performanceComparison}
\begin{tabular}{|l|c| c|c|}
\hline
{\bf Experiments by}&\mc{3}{|c|}{\bf Attacks}\\
\hline
&{\bf Launched}&{\bf Detected (\waterdefense)}&{\bf Detected (\waterdefenseH})\\
\hline
Authors&37&33 (89\%)&NA\\
\hackfest-2016 &18&10 (55.5\%)&NA\\
\hackfest-2017&31&21 (67.7\%)&24 (77.4\%)\\
\hline
\mc{4}{p{\linewidth}}{NA: \waterdefenseH did not exist at the time of experimentation by the author and during \hackfest-2016.}

\end{tabular}
\end{table}

The data in Table~\ref{tab:performanceComparison} is indicative of the value of organizing \hackfest events. Specifically, in the case described in this paper, the two \hackfest events led to an increased confidence in the effectiveness of the invariant-based approach in detecting cyber attacks. The hackfests also led to the creation of new types of attack vectors that were not used earlier to assess the performance of \waterdefense and \waterdefenseH in detecting cyber attacks.

\subsection{False alarms}

The performance of any attack detection method ought to be assessed using its detection accuracy, i.e. how many of the launched attacks it detects, as well as the rate at which false alarms are raised.  During \hackfest each team  attempted to launch several attacks. The attacks listed in Tables~\ref{tab:2016attacks} and \ref{tab:2017attacks} are the ones that were successful in realizing the stated attacker intent and were scored by the judges. The remaining attacks were not recorded and hence any alarm generated by such attacks was not considered. Some of these unrecorded alarms could be false though no specific claims can be made about their nature. 

Since \hackfest-2017, the  authors have observed no false alarms  from \waterdefense during normal operation of SWaT.  \waterdefenseH has been in operation since a few weeks  prior to \hackfest-2017. Again, during the normal operation of SWaT, no alarm has been generated by \waterdefenseH. This observation should not be construed to imply  that an invariant-based attack detection mechanism will not generate any false alarm-- in fact it could. However, if the invariants generated are  complete in the sense that they accurately capture {\em all} aspects of process behavior, and their implementation is correct and tuned properly, the likelihood of false alarms is low.  

Even though SWaT is a relatively new plant (2-years since its inauguration at the time of writing this paper), we do observe intermittent failures in a few motorized valves. For example, sometimes MV101 in Stage~1 takes much longer  to open than expected by its controlling PLC1. The PLC itself detects such cases. In such a case \waterdefense or \waterdefenseH, depending on the time it takes for the valve to finally open, will raise an alarm. We do not consider this as a false positive simply because whether an anomalous behavior is due to a natural cause, or a cyber attack, cannot be distinguished by \waterdefense or \waterdefenseH. While such distinction is important to make, additional research is needed to distinguish process anomalies due to cyber attacks and those arising due to natural component failures.

\subsection{Benefits of \hackfest}
\hackfest exposed the  organisers, participants and researchers to how an attacker might design and launch attacks on ICS. Benefits of \hackfest include the following. 1)\,An improved understanding of how an ICS operates and the consequent formulation of new research directions. 2)\, Opportunity for participants from industry and academia to learn from the event and focus on the limitations of their work. 3)\,An aid to the ICS management team to observe the defense teams thus leading to possible  adoption of technology embedded in \waterdefense or \waterdefenseH.

\subsection{Placement of \waterdefense}

The placement of \waterdefense  is another question that ought to be looked into carefully. In this work \waterdefense  is placed inside PLCs. However, an exceptionally large number of invariants may prevent adding code to the existing control code in a PLC. This may happen due to the computational load requirements on a PLC. This aspect led us to create \waterdefenseH that is placed on the plant network and gets its data from the Historian to evaluate the invariants. 

\subsection{Forensics}

One  advantage of the invariant-based approach for attack detection appears while determining the area of impact of an attack.   When a single invariant is violated, it indicates clearly  the source of process anomaly. For example, an alert is generated if valve MV101 is closed when the water in tank T101 is at or below the \state{L} level marker. While this alert does not indicate how an attacker entered the system, or if the valve or the level sensor is defective, it does assist in localising the reason for the alert.  The analysis becomes a bit more complex when multiple invariants raise alerts. This aspect of an invariant-based detection mechanisms remains to be analyzed in further detail.


\subsection{Attacker capabilities}

 We do not have any validation of the professionalism of the \hackfest attack teams. As mentioned earlier\,\cite{s3:2017Report,s3-16,s3-17}, attack teams were from a variety of backgrounds including from the industry and  academia from Europe and Asia.  During \hackfest-2017 one team consisting of four members--all from outside of Singapore-- focuses on ethical hacking and cyber-wars involving critical infrastructure. This team is part of a global alliance. The other teams consist of hackers interested in knowing how vulnerabilities in software can be exploited and passes this information to others for improving systems security. Coverage of attacks launched by the attack teams, and attacker profiles, is discussed in Section~\ref{sec:eventDetails}  and summarized in Tables~\ref{tab:attacker_profiles}, \ref{tab:attackTargets}, \ref{tab:cybercriminal-attacks} and \ref{tab:2017results}.
 
\subsection{Attack trees}

It is possible to use attack trees\,\cite{vineetQiangVamsi,tenLiuGovindrasu} to model  attacks launched during  the two hackfests reported in this paper. Doing so would enable mapping each attack to a specific path in the attack tree and reveal which attack paths in SWaT were traversed. Such modeling and analysis has not been attempted in this work and is a possible subject for future research.

\section{Related Work}
\label{sec:relatedwork}

\hackfest is a Capture-The-Flag\,\cite{ctftime} event on ICS. Traditional CTF events generally attract the attention of both industrial and academic teams and currently enjoy increasing popularity as indicated in\,\cite{ctftime}. The number of such events is   gradually increasing\,\cite{childers2010,defcon}. Such  events aid in learning about security vulnerabilities, how these could be exploited, nature of attacks, and strength of the deployed \,\cite{eagle04capture,radcliffe07capture,vigna2003teaching} defense mechanisms.    
To the best of our knowledge, \hackfest is the first CTF style event  of its kind in ICS  that involves participants from the industry  and  academia, and   focuses on  an operational  water treatment testbed. 

The study reported here focuses on cyber attacks on ICS that result in deliberate  data and command manipulation. Injection of such  attacks in ICS  has been studied by several researchers. Attacks have been modeled as noise in sensor data\,\cite{kwonLiuHwang2013,weerakkodyMoSinopoli}. Authors previously presented cyber physical attacker model\,\cite{adepuAttackerModel}  to aid in the design of cyber physical attacks on ICS. Attacker models designed specifically for ICS include a variety of deception attacks including surge, bias, and geometric\,\cite{cardenasAminLinHuangHuangSastry}. Such models have been used in experiments to understand the effectiveness of statistical techniques in detecting cyber attacks. 

There exist several techniques, other than the type used in \waterdefense, for  the detection of process anomalies.  CPAC\,\cite{etigowni2016cpac} presents stateful detection mechanisms to detect attacks against control systems. The Weaselboard\,\cite{mulder2013weaselboard}  uses PLC backplane to get the sensor data and actuator commands, and analyses them to prevent zero day vulnerabilities. WeaselBoard\,\cite{mulder2013weaselboard}  has a dedicated device, and detects changes in control settings, sensor values, configuration information, firmware, logic, etc. 

 The invariants in \waterdefense  use data from multiple stages to enable   distributed detection  of cyber attacks. Such sensor fusion has been proposed by several researchers.  In safety critical cyber physical systems this was reported in\,\cite{ivanov2016attack}. In\,\cite{Stankovic2016}, it is shown how  safety critical systems are  interconnected and their complexity. Model based attack detection schemes in water distribution systems was presented in\,\cite{chuadhry2017model}. It uses the Matlab system identification tool to get a model from the data generated in a water distribution system. The data driven model is helpful in detecting process anomalies. 

Monitoring the physics of the system has been studied in\,\cite{gollmanKrotofil}. Cardenas et al.\,\cite{urbinaGiraldoCardenasTippenhauerValenteFaisalRuthsCandellSandberg}  have experimented with the use of  CUSUM  in detecting  stealthy attacks.
Hsio et al.\,\cite{hsiao2010cross} have proposed a distributed security monitoring solution to detect attacks on an ICS. There exists literature on the design of robust ICS\,\cite{kwonLiuHwang2013,wasicekDerlerLee}. These works focus on attack modelling and the design of controllers and monitors for secure ICS.

\section{Conclusion}
\label{sec:conclusion}

There exist a number of devices for defending networks and ICS against cyber attacks. Firewalls attempt to prevent attackers from entering an ICS. Intrusion Detection Systems (IDSs) attempt to detect if an unauthorized user has entered the plant network.  The approach used in WD is orthogonal to that used in most commercially available firewalls and IDS. \waterdefense  uses a design-centric approach to detect process anomalies in contrast to network traffic anomalies that are the focus of several IDS. Thus, \waterdefense is effective in detecting attacks by an external or an internal agent. One could consider WD as a last-mile defense.

While in the study reported here WD has been found effective in detecting attacks that lead to process anomaly, it does fail in detecting attacks such as a replay attack where a plant operator views the system state that is different from the actual state. This ineffectiveness of WD ought to be considered when using such a system in critical infrastructure. 

It is interesting to observe that there exist attacks that are detected by both \waterdefense and \waterdefenseH though  vice-versa is not true. For example,   
attack\,\ref{Historian_aircrack} in Table~\ref{tab:2017attacks} was detected by \waterdefenseH but not by \waterdefense. This observation suggests that, when feasible, both systems ought to be deployed simultaneously. 

The invariants used in \waterdefense and \waterdefenseH were derived and coded manually. For a system such as SWaT the manual approach is feasible as the plant has 42 sensors and actuators as compared to perhaps hundreds or more in commercial plants.  Thus, there needs to be an automated way of generating and coding the invariants. 

The attacks launched by teams during the hackfests could later serve as a source for assessing the effectiveness of attack detection mechanisms developed by other researchers. Details of all attacks launched during the hackfests are therefore made public and available in\,\cite{antonioliGhaeiniAdepuOchoaTippenhauer,s3:2017Report,itrustDataset}.

It should be obvious that any attack detection mechanism, including \waterdefense, is one component of a holistic defense system against cyber attacks on any critical infrastructure. This paper does not address an important question: {\em What action should be taken, and how, when an alarm is raised by \waterdefense or \waterdefenseH ?"} This remains an open question.

\section*{Acknowledgments}
\noindent  A  number of people were involved in the planning, execution, and post-data analysis during the two hackfests reported in this paper. Our  thanks are due to Nils Tippenhauer, Martin Ochoa, and the staff of iTrust for organizing and judging the events; Kaung Myat Aung  for invaluable assistance in the actual conduct of the events;  Gyanendra Mishra for implementing \waterdefenseH; the entire team of authors of the \hackfest-2017 report\,\cite{s3:2017Report}, namely Francisco Furtado, 
Lauren Goh, Sita Rajgopal, Elaine Cheung, Ericson Thiang, Toh Jing Hui, and Ivan Lee; to the SUTD-MIT International Design Center for partially supporting \hackfest-2017; and to all the participants who traveled long distances to come to Singapore to participate in the two hackfests. Last but not the least, thanks to the reviewers for their comments that helped improve the original manuscript.

\bibliographystyle{abbrv}

\bibliography{Security.bib}

\section*{Biography}
\skipnoindent {\bf  Sridhar Adepu} is a PhD student in Information Systems Technology and Design pillar at the  Singapore University of Technology and Design. His research focuses on verification, safety, security and reliability of Cyber-Physical Systems. 

\skipnoindent {\bf  Aditya Mathur} is a Professor  of Computer Science at Purdue University and Head of Pillar Information Systems Technology and Design at the Singapore University of Technology and Design. Aditya is Center Director of iTrust, a center for research in cyber security.  Design of secure public infrastructure is a focus of his current research.


\end{document}